\documentclass[sigconf, authorversion]{acmart}

\usepackage{booktabs} 
\setcopyright{rightsretained}

\usepackage{array}
\usepackage{color}
\usepackage[utf8]{inputenc}
\usepackage{dblfloatfix}
\usepackage{cleveref}

\definecolor{orange}{rgb} {0.9,0.5,0.0}
\definecolor{green}{rgb} {0,0.5,0.3}

\begin{document}
\title{Towards Artificial Learning Companions for Mental Imagery-based Brain-Computer Interfaces}

\author{L.\, Pillette}
\affiliation{%
  \institution{Inria, LaBRI (Univ. Bordeaux, \\CNRS, Bordeaux-INP), France}}
\email{lea.pillette@inria.fr}

\author{C.\, Jeunet}
\affiliation{%
  \institution{Univ. Rennes, Inria, IRISA, CNRS, France / CNBI, EPFL, Switzerland}}
  \email{camille.jeunet@inria.fr}

\author{R.\, N'Kambou}
\affiliation{%
  \institution{GDAC, UQAM, Quebec}
  \country{Canada}}
  \email{nkambou@gmail.com}
  
\author{B.\, N'Kaoua}
\affiliation{%
  \institution{Handicap, Activity, Cognition, Health, Univ. Bordeaux, CNRS, France}}
  \email{bernard.nkaoua@u-bordeaux.fr}
  
\author{F.\, Lotte}
\affiliation{%
  \institution{Inria, LaBRI (Univ. Bordeaux, \\CNRS, Bordeaux-INP), France}}
\email{fabien.lotte@inria.fr}
  
\author{}

\author{}

\begin{abstract}
Mental Imagery based Brain-Computer Interfaces (MI-BCI) enable their users to control an interface, e.g., a prosthesis, by performing mental imagery tasks only, such as imagining a right arm movement while their brain activity is measured and processed by the system. Designing and using a BCI requires users to learn how to produce different and stable patterns of brain activity for each of the mental imagery tasks. However, current training protocols do not enable every user to acquire the skills required to use BCIs. These training protocols are most likely one of the main reasons why BCIs remain not reliable enough for wider applications outside research laboratories. Learning companions have been shown to improve training in different disciplines, but they have barely been explored for BCIs so far.
This article aims at investigating the potential benefits learning companions could bring to BCI training by improving the feedback, i.e., the information provided to the user, which is primordial to the learning process and yet have proven both theoretically and practically inadequate in BCI. This paper first presents the potentials of BCI and the limitations of current training approaches. Then, it reviews both the BCI and learning companion literature regarding three main characteristics of feedback: its appearance, its social and emotional components and its cognitive component. From these considerations, this paper draws some guidelines, identify open challenges and suggests potential solutions to design and use learning companions for BCIs.

\section*{RESUME}
Les interfaces cerveau-ordinateur (ICO) exploitant l'imagerie mentale permettent à leurs utilisateurs d’envoyer des commandes à une interface, une prothèse par exemple, uniquement en réalisant des tâches d’imagerie mentale, tel qu'imaginer son bras droit bouger. Lors de la réalisation de ces tâches, l'activité cérébrale des utilisateurs est enregistrée et analysée par le système. Afin de pouvoir utiliser ces interfaces, les utilisateurs doivent apprendre à produire différents motifs d’activité cérébrale stables pour chacune des tâches d'imagerie mentale. Toutefois, les protocoles d'entraînement existants ne permettent pas à tous les utilisateurs de maîtriser les compétences nécessaires à l’utilisation des ICO. Ces protocoles d'entraînements font très probablement partie des raisons principales pour lesquelles les ICO manquent de fiabilité et ne sont pas plus utilisées en dehors des laboratoires de recherche. Or, les compagnons d’apprentissage, qui ont déjà permis d’améliorer l'efficacité d'apprentissage pour différentes disciplines, sont encore à peine étudiés pour les ICO. L’objectif de cet article est donc d’explorer les différents avantages qu’ils pourraient apporter à l'entraînement aux ICO en améliorant le retour fait à l'utilisateur, c'est-à-dire les informations fournies concernant la tâche. Ces dernières sont primordiales à l'apprentissage et pourtant, il a été montré qu'à la fois théoriquement et en pratique ces dernières étaient inadéquates. Tout d'abord, seront présentés dans l'article les potentiels des ICO et les limitations des protocoles d'entraînement actuels. Puis, une revue de la littérature des ICO ainsi que des compagnons d'apprentissage est réalisée concernant trois caractéristiques principales du retour utilisateur, c'est-à-dire son apparence, ses composantes sociale et émotionnelle et enfin sa composante cognitive. À partir de ces considérations, ce papier fournit quelques recommandations, identifie des défis à relever et suggère des solutions potentielles pour concevoir et utiliser des compagnons d'apprentissage en ICO.
\end{abstract}


\maketitle

\section{INTRODUCTION}
A Brain Computer Interface (BCI) can be defined as a technology that enables its users to interact with computer applications and machines by using their brain activity alone \cite{Clerc16-v1}. In most BCIs, brain activity is measured using Electroencephalography (EEG), which uses electrodes placed on the scalp to record small electrical currents reflecting the activity of large populations of neurons \cite{Clerc16-v1}. In a BCI, EEG signals are processed and classified, in order to assign a specific command to a specific EEG pattern. For instance, a typical BCI system can enable a user to move a cursor to the left or right on a computer screen, by imagining left or right hand movements, each imagined movement leading to a specific EEG pattern \cite{Pfurtscheller01}. In this article we focus on Mental Imagery-based BCI (i.e., MI-BCI) with which users have to consciously modify their brain activity by performing mental imagery tasks (e.g., imagining hand movements or mental calculations) \cite{Clerc16-v1,Pfurtscheller01}. MI-BCIs require the users to train to adapt their own strategies to perform the mental imagery task based on the feedback they are provided with. At the end of the training, the system should recognize which task the user is performing as accurately as possible. However, it has been shown, both theoretically and practically, that the existing training protocols do not provide an adequate feedback for acquiring these BCI skills \cite{Lotte13, Jeunet16}. This, among other reasons, could explain why BCIs still lack reliability and that around 10 to 30\% of users cannot use them at all \cite{LotteHDR2016, Neuper10}. Several experiments showed that taking into account recommendations from the educational psychology field, e.g., providing a multisensorial feedback, can improve BCI performances and user-experience \cite{Sollfrank16, Lecuyer08}. However, researches using a social and emotional feedback remain scarce despite the fact that it is recommended by educational psychology \cite{Goleman95}. 

Indeed, it has been hypothesized that our social behavior had a major influence in the development of our brain and cognitive abilities \cite{Dunbar07, Ybarra08}. Social interaction was traditionally involved in the intergenerational transmission of practices and knowledge. However, its importance for learning was acknowledged only recently with the development of the social interdependence theory, which states that the achievement of one person's goal, i.e., here learning, depends on the action of others. Cooperative learning builds on this idea and promotes collaboration between students in order to reach their common goal \cite{Johnson09}. These theories and methods have shown that learning can be strengthened by a social feedback \cite{Izuma08, Chou03}. 

Artificial learning companions, which are animated conversational agents involved in an interaction with the user \cite{Chou03}, could provide such social and emotional feedback. Physiological and neurophysiological data recordings offer the possibility to infer users' states/traits and to adapt the behavior of the companion accordingly \cite{Burleson07, Bent17}. The training would benefit from the later, for example the difficulty of the task could be modulated in order to keep the user motivated. In particular, the feedback provided during the training could be improved, e.g., by adapting to the emotional state of the user. Learning companions are therefore able to take into account the cognitive abilities and affective states of users, and to provide them with emotional or cognitive support. They have already proven to be effective for improving learning of different abilities, e.g., mathematics or informatics, \cite{Cabada12, Kim05}. From all types of computational supports which enrich the social context during learning (i.e., educational agent) we chose to focus on learning companions because they engage in a non-authoritative interaction with the user, can have several roles ranging from collaborator, to competitor or teachable student and could potentially involve using several of them with complementary roles \cite{Chou03}.

Learning companions could contribute to improving BCI training by, among other, enriching the social context of BCI. This articles aims at identifying the various benefits that learning companions can bring to BCI training, and how they can do so. To achieve this objective, this article starts by detailing the principles and applications of BCI as well as the limitations of current BCI training protocols. Once the keys to understanding BCIs provided, this article focuses on three main components of BCI feedback, which should be improved in order to improve BCI training. First of all, we study the appearance of feedback, which is one of its most studied characteristics. Second, we study its social component, i.e., the amount of interaction the user has with a social entity during the learning task, and emotional component, i.e., the  feedback components which aim at eliciting an emotional response from the user. Both are still scarcely used in BCI though the existing results seem to be promising. Finally, we will concentrate on its cognitive component, i.e., which information to provide users in order to improve their understanding of the task, which represents one of the main challenge in designing BCI feedbacks. For each of these three feedback components, we propose a review of the literature for both the BCI and learning companion fields to deduce from them some guidelines, challenges and potential research directions.

\section{BRAIN COMPUTER INTERFACE SKILLS}

\subsection{BCIs principes and applications}

Since they make computer control possible without any physical movement, MI-BCIs rapidly became promising for a number of applications \cite{Clerc16-v2}. They can notably be used by severely motor impaired users, to control various assisting technologies such as prostheses or wheelchairs \cite{Millan10}. More recently, MI-BCIs were shown to be promising for stroke rehabilitation as well, as they can be used to guide stroke patients to stimulate their own brain plasticity towards recovery \cite{Ang15}. Finally, MI-BCIs can also be used beyond medical applications \cite{vanErp12}, for instance for gaming, multimedia or hand-free control, among many other possible applications \cite{Clerc16-v2}. However, as it has been mentioned, despite these many promising applications, current EEG-based MI-BCIs are unfortunately not really usable, i.e., they are not reliable nor efficient enough \cite{LotteHDR2016,Clerc16-v1,Clerc16-v2}. In particular, the mental commands from the users are too often incorrectly recognized by the MI-BCI. There is thus a pressing need to make them more usable, so that they can deliver their promises.

Controlling a MI-BCI is a skill that needs to be learned and refined: the more users practice, the better they become at MI-BCI control, i.e., their mental commands are recognized correctly by the system increasingly more often \cite{Jeunet16e}. Learning to control an MI-BCI is made possible thanks to the use of neurofeedback (NF) \cite{Sitaram16}. NF consists in showing users a feedback on their brain activity, and/or as with BCI, in showing them which mental command was recognized by the BCI, and how well so. This is typically achieved using a visual feedback, e.g., a gauge displayed on screen, reflecting the output of the machine learning algorithm used to recognize the mental commands from EEG signals \cite{Neuper10} (see Figure \ref{fig:feedback}). This guides users to learn to perform the MI tasks increasingly better, so that they are correctly recognized by the BCI. Thus, human learning principles need to be considered in BCI training procedures \cite{Lotte15a}.

\subsection{Limitations of the current training protocol} 

\begin{figure}[htb!]
\begin{center}
\includegraphics[width=8cm]{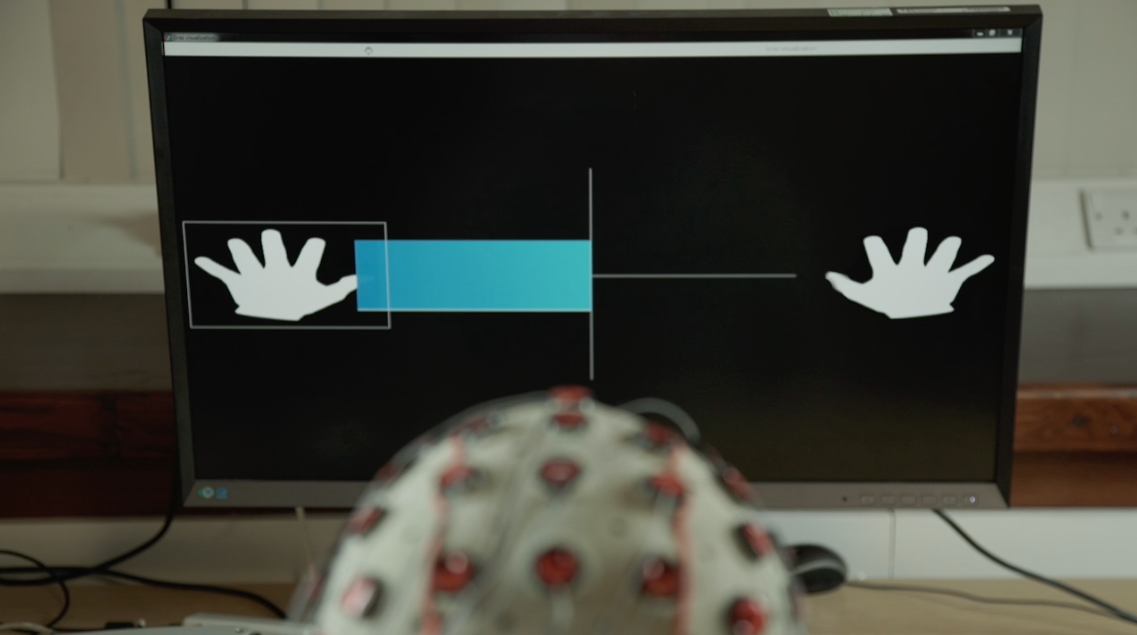}
\caption{Example of feedback which is often provided to users during training, i.e., right and left hand motor imagery training here. At the moment the picture was taken the user had to imagine moving his left-hand. The blue bar indicates which task has been recognized and how confident the system is in its recognition. The longer the bar and the most confident the system is. Here the system rightly recognize the task that the user is performing and is quite confident about it \cite{Pfurtscheller01}.}
\label{fig:feedback}
\end{center}
\end{figure}

Currently, most MI-BCI studies are based on the Graz training protocol or on variants of the latter. This protocol relies on a two stage procedure \cite{Pfurtscheller01}: (1) training the system and (2) training the user. In stage 1, the user is instructed to successively perform a certain series of MI tasks (for example, left and right hand MI). Using the recordings of brain activity generated as these various MI tasks are performed, the system attempts to extract characteristic patterns of each of the mental tasks. These extracted features are used to train a classifier, the goal of which is to determine the class to which the signals belong. 
Then, in stage 2 users are instructed to perform the MI tasks, but this time feedback (based on the system training performed in stage 1) is provided to inform them of the MI task recognized by the system. The user’s goal is to develop effective strategies that will allow the system to easily recognize the MI tasks that they are performing. Along such training, participants are asked to perform specific mental tasks repeatedly, e.g., imagining left or right-hand motor imagery, and are provided with a visual feedback shown as a bar indicating the recognized task and the corresponding confidence level of the classifier (see Figure \ref{fig:feedback}).
Unfortunately, such standard training approaches satisfy very few of the guidelines from human learning psychology and instructional design to ensure an efficient skill acquisition \cite{Lotte13}. For instance, a typical BCI training session provides a uni-modal (visual) and corrective feedback (indicating whether the learner performed the task correctly) (see Figure \ref{fig:feedback}), using fixed and (reported as) boring training tasks identically repeated until the user achieves a certain level of performance, with these training tasks being provided synchronously. In contrast, it is recommended to provide a multi-modal and explanatory feedback (indicating what was right/wrong about the task performed by the user) that is goal-oriented (indicating a gap between the current performance and the desired level of performance), in an engaging and challenging environment, using varied training tasks with adaptive difficulty \cite{Merrill07, Shute08}. 

Moreover, it is necessary to consider users' motivational and cognitive states to ensure they can perform and learn efficiently \cite{Keller08}. Keller states that optimizing motivational factors - Attention (triggering a person’s curiosity), Relevance (the compliance with a person’s motives or values), Confidence (the expectancy for success), and Satisfaction (by intrinsic and extrinsic rewards) - leads to more user efforts towards the task and thus to better performance. 

In short, current standard BCI training approaches are both theoretically \cite{Lotte13} and practically \cite{Jeunet16} suboptimal, and are unlikely to enable efficient learning of BCI-related skills. Artificial intelligent agents such as learning companions could provide tools to improve several characteristics of BCI training.

\section{BUILDING BCI LEARNING COMPANION - EXISTING TOOLS AND CHALLENGES}
Learning companions have been defined by \cite{Chou03} as follows:
\begin{quote}In an extensive definition, a learning companion is a computer-simulated character, which has \textbf{human-like characteristics} and plays a \textbf{non-authoritative role} in a \textbf{social learning environment}.\end{quote}
This definition offers three main points that will be elaborated in the BCI context in the following section.
First, the learning companion must facilitate the learning process in particular by encouraging the learner in a social learning activity. Using an anthropomorphic appearance facilitates this social context. Furthermore, its interventions should be consistent with the general recommendation concerning feedback which would also contribute to its human likeness and its efficiency \textit{(See Section \ref{chap:AspectFB})}.

Second, learning companions are educational agents, i.e., computational supports which enrich the social context during learning \cite{Chou03}. Such environment could provide a motivating and engaging context that would favor learning \textit{(See Section \ref{chap:SocialEmotionalFB})}.

Finally, the benefit of a learning companion over the other types of educational agents is that its role can greatly vary from student to tutor given the learning model used and the knowledge that the companion holds. At the moment, an educational agent with an authoritative role of teacher is not realistic because of the lack of a cognitive model of the task. Such a model would provide information about how the learner’s profile (i.e., traits and states) influences BCI performance and which feedback to provide accordingly \cite{Jeunet16, Jeunet17}. It would be necessary to understand, predict and therefore improve the acquisition of BCI skills \textit{(See Section \ref{chap:CognitiveFB})}. 


\subsection{Appearance of feedback}\label{chap:AspectFB}
As stated above, the appearance of the learning companion greatly impacts its influence on the user. BCI performances are also influenced by the appearance of the feedback that is provided during training. Therefore, numerous researches have been and are still led toward improving this characteristic of the feedback.

\subsubsection{BCI Literature}
While it is recognized that feedback improves learning, many authors have attempted to clarify which features enhance this effect \cite{Azevedo95, Bangert91, Narciss04}. 
To be effective, feedback should be directive (indicating what needs to be revised), facilitative (providing suggestions to guide learners) and should offer verification (specifying if the answer is correct or incorrect).
It should also be goal-directed by providing information on the progress of the task with regard to the goal to be achieved. Finally, feedback should be specific, clear, purposeful and meaningful. These different features increase the motivation and the engagement of learners \cite{Williams97, Hattie07, Ryan00}.

As already underlined in \cite{Lotte13}, classical BCI feedback satisfies few of such requirements. Generally, BCI feedbacks are not explanatory (they do not explain what was good or bad nor why it is so), nor goal directed and do not provide details about how to improve the answer. Moreover, they are often unclear and do not have any intrinsic meaning to the learner. For example, BCI feedback is often a bar representing the output of the classifier, which is a concept most BCI users are unfamiliar with.

Recently, some promising areas of research have been investigated. For example, the study in \cite{Kubler01a} showed that performances are enhanced when feedback is adapted to the characteristics of the learners. In their study, positive feedback, i.e., feedback provided only for a correct response, was beneficial for new or inexperienced BCI users, but harmful for advanced BCI users. 

Several studies also focused on the modalities of feedback presentation. The work in \cite{Ramos12} used BCI for motor neurorehabilitation and observed that proprioceptive feedback (feeling and seeing hand movements) improved BCI performance significantly. In a recent study in \cite{Jeunet15a}, the authors tested a continuous tactile feedback by comparing it to an equivalent visual feedback. Performance was higher with tactile feedback indicating that this modality can be a promising way to enhance BCI performances. 
The study in \cite{Sollfrank16} showed that multimodal (visual and auditory) continuous feedback was associated with better performance and less frustration compared to the conventional bar feedback.

\begin{center}
\begin{figure}[htb!]
\includegraphics[width=7.5cm]{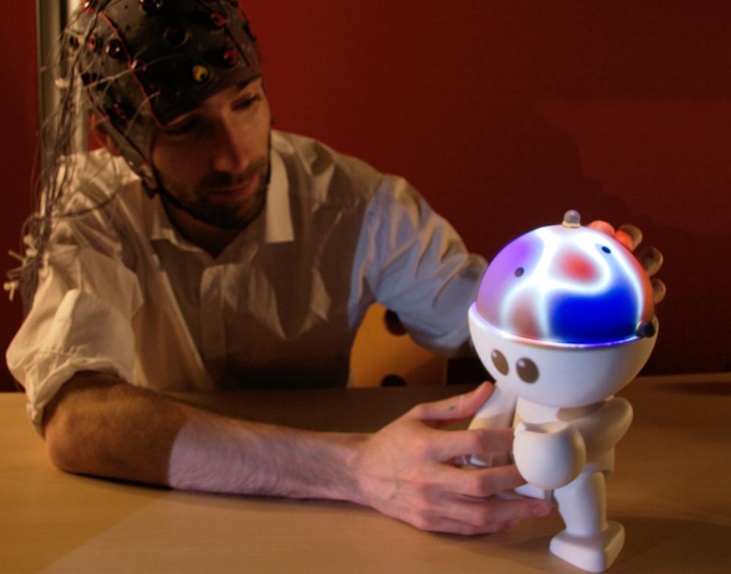}
\caption{User visualizing his brain activity using Teegi \cite{Frey14}.}
\label{fig:teegi}
\end{figure}
\end{center}

Other studies investigated new ways of providing some task specific and more tangible feedback. In \cite{Frey14} and \cite{Mercier14}, the authors created tools using augmented reality to display the user's EEG activity on the head of a tangible humanoid called Teegi (see Figure \ref{fig:teegi}), and superimposed on the reflection of the user, respectively. 
These researches contribute to make feedback more attractive which can have a beneficial impact. For example, it has been shown that using game-like, 3D or Virtual Reality increase user engagement and motivation \cite{RonAngevin09}. 

\subsubsection{Learning Companion Literature}
Numerous researches regarding the appearance which would maximize the acquisition of a skill/ability were led as well for learning companions. Some main points seem to emerge from them, here are some that could be useful while designing one companion for BCI purpose:
\begin{itemize}
\item Physical, tangible companion seems to increase social presence in comparison to a virtual companion \cite{Hornecker11, Schmitz10}
\item Anthropomorphic features facilitate social interactions \cite{Duffy03}
\item Physical characteristics, personality/abilities, functionalities and learning function should be consistent \cite{Norman94}
\end{itemize}
Interestingly, the influence of learning companions was also studied using measures of brain activity. For example, using functional Magnetic Resonance Imaging (fMRI) \cite{Krach08} investigated the neural correlates of the attribution of intentions and desires (i.e., theory of mind) for different robot features. Results show that theory of mind-related cortical activity is positively correlated to the perceived human-likeness of a robot. This implies that the more realistic the robots, the more people attribute them intentions and desires. 

\subsubsection{Futures challenges} 
Feedbacks that are both adapted and adaptive to users are lacking in both the BCI and learning companion literatures. Numerous researches have been and are still led toward identifying feedback characteristics and learner characteristics influencing BCI performance. However, the often low number of participants in current experiments limits those results and further researches should be led to clarify the type of feedback to provide depending on the user's profile. 

Additionally, an interesting research direction could be to use several learning companions, including Teegi or another tangible system which could display the brain activity of the user. Each companion could have a different role and one of them could be a tutor which would provide insights about how to interpret the information related to brain activity displayed.

\subsection{Social \& Emotional feedback} \label{chap:SocialEmotionalFB}
Learning companions are more than just another mean to provide feedback. Their main benefits is that they enrich the social context of learning and can provide emotional feedback. As mentioned, BCI training still lacks such elements in its feedback though current literature tends to indicate that it would benefit from them.

\subsubsection{BCI Literature}
Indeed, \cite{Nijboer08} showed that mood, assessed prior to each BCI session (using a quality of life questionnaire), correlates with BCI performances. Some BCI experiments provided emotional feedback using smiling faces to indicate the user if the task performed had been recognized by the system \cite{Kubler01b, Leeb07}. Though, none of these studies used a control group. Therefore, the impact of such a feedback remains unknown for BCI applications. 
A similar study was led in neurofeedback by \cite{Mathiak15} who showed that providing participants with an emotional and social feedback as a reward enabled better control than a typical moving bar over the activation of the dorsal anterior cingulate cortex (ACC) monitored using fMRI. The feedback consisted of an avatar’s smile whose width varied depending on the user’s performance. The better the performance, the wider the smile was. This type of feedback can be considered as both emotional and social because of the use of an avatar. 

The use of social feedback in BCI has been encouraged in several papers \cite{Sexton15, Lotte13, Mattout12}. The work in \cite{Izuma08} showed that a social feedback can be considered as a reward just as much as a monetary one. Yet, the influence of a reward has already been demonstrated in BCI. Indeed, it has been shown that a monetary reward can modulate the amplitude of some brain activity, including the one involved during MI-BCI \cite{Sepulveda16, Kleih10}. However, researches about the use of a social feedback in BCI remain scarce and often lack of control groups. One of the main original purpose of BCI was to enable their users to communicate and some researchers have created tools to provide such type of communication in social environments, for example using Twitter \cite{Edlinger11} but no comparison was made with equivalent nonsocial environment. 
Studies from \cite{Bonnet13}, \cite{Obbink12} and \cite{Goebel04} presented games where users played in pairs collaborating and/or competing against each other. The study in \cite{Bonnet13} found that this type of learning context proved successful to significantly improve user-experience and the performances of the best performing users. 

\begin{center}
\begin{figure}[htb!]
\includegraphics[width=8.5cm]{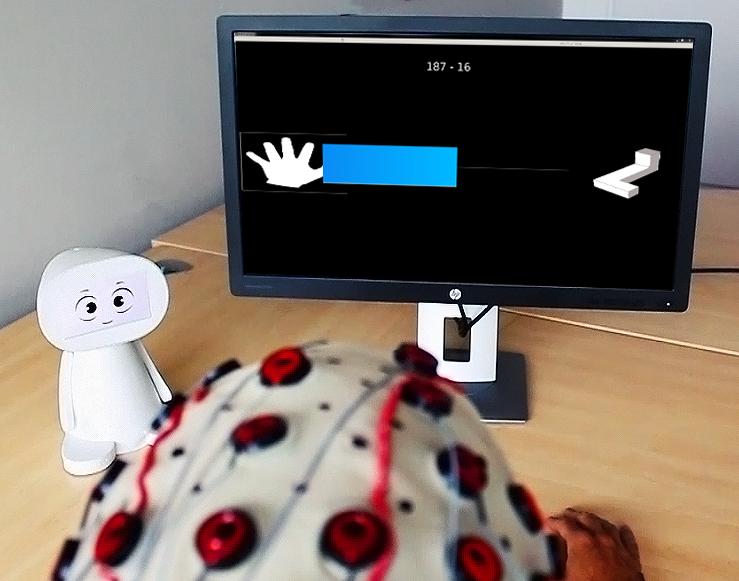}
\caption{Experimental setting where PEANUT (on the left) provides a user with social presence and emotional support adapted to his performance and progression \cite{Pillette17}.}
\label{fig:peanut}
\end{figure}
\end{center}

Finally, we explored the use of social and emotional feedback when creating PEANUT (i.e., Personalized Emotional Agent for Neurotechnology User Training), which is the first learning companion dedicated to BCI training \cite{Pillette17}. Its interventions were composed of spoken sentences and displayed facial expression in between two trials (see Figure \ref{fig:peanut}). The interventions were selected based on the performance and progression of the user. We tested PEANUT’s influence on user’s performance and experience during BCI training using two groups of users with similar profiles. One group was trained to use BCI with PEANUT and the other without. Our results indicated that the user experience was improved when using PEANUT. Indeed, users felt that they were learning and memorizing better when they were learning with PEANUT. Even though their mean performances were not changed, the variability of the performances from the group with PEANUT were significantly higher than for the other group. Such result might indicate a differential effect of learning companions on users \cite{Burleson07}.

\subsubsection{Learning Companion Literature}
Several other studies have shown the interest of learning-companions as source of social link that is sometimes essential in certain learning situations \cite{Lester97,Saerbeck10}. They can play different roles such as co-learner, co-tutor, etc. in which they are often called upon to demonstrate certain capacities of social interaction such as empathy through emotional feedback \cite{Lester97} or respect for social norms \cite{Johnson04, Saerbeck10}.

Emotional feedback aims at regulating the emotions of the learner throughout the learning. Positive emotions are known to improve problem solving, decision-making, and creation, while negative emotions are harmful in these situations \cite{Isen01}. Previous studies regarding emotional feedback investigated emotional regulation strategies to manage learners’ emotions and behaviors \cite{Beale09, Burleson07, Mcquiggan10}. The positive impact of emotional feedback has also been highlighted in some educational contexts \cite{Terzis12}.

In addition, it is important to adapt the social interaction to each learner. Indeed, it has been shown that a companion that adapts its behavior to learners' profile increases the development of their positive attitude \cite{Gordon16}. 

Learning companions are sometimes embodied in robots to better materialize social presence. Tega is a social companion robot which interprets students' emotional response -- measured from facial expressions -- in a game aimed at learning Spanish vocabulary \cite{Gordon16}. It approximates the emotions of the learner and over time, determines the impact of these emotions on the learner to finally create a personalized motivational strategy adapted to the later.

To ensure adaptation, machine learning techniques are often deployed. With the advancement of Artificial Intelligence (AI), more efficient techniques are now used to help the companion to better learn from the learner's behavior. In case of the social companion NICO (a Neuro-Inspired COmpanion robot), the model used for the learning of the emotions and the adaptation to the user is a combination of a Convolutional Neural Network and a Self-organization Map to recognize an emotion from the user's facial expression, and learn to express the same \cite{Churamani17}. The model allows the robot to adapt to a different user by associating the perceived emotion with an appropriate expression which makes the companion more socially acceptable in the environment in which it operates.
\subsubsection{Futures challenges}
As mentioned above, assessing users' emotional states is particularly useful for learning companions. However, doing so reliably remains a challenge, particularly for covert emotions, that are thus not visible in facial expressions. Passive BCI, with which the brain activity is analyzed to provide metrics related to the mental state of the user to adapt the system accordingly could be used for this purpose \cite{Zander09}. Though monitoring emotional states remains a challenge because experiments compare their emotional recognition performances using self reported measures from the users. Such experimental protocol assumes that people are able to self assess reliably their own emotions which might not be true \cite{Robinson02}. Furthermore, the brain structures involved in emotional states are sub-cortical, e.g., amygdala \cite{LeDoux95}, which means that reliably monitoring them using non invasive EEG is an issue \cite{Muhl14}. Nevertheless, some promising results were found in particular using EEG \cite{Muhl14} (see Table \ref{tab:summary}).

It also remains to be evaluated whether using a learning companion can help to reduce the fear induced by the BCI setup that has a detrimental effect on BCI performances \cite{Witte13}.

\begin{table*}[bp]
\begin{tabular} { | m{2.4cm} | m{4.6cm} | m{4.6cm}| m{4.6cm} | }
 \hline
  & 
  Recommendations & 
  Challenges & 
  Potential solutions \\ 
 \hline
  \hline
 Appearance of Feedback & 
 BCI feedback should take into consideration of recommendations from educational psychology, e.g., multisensorial \cite{Sollfrank16} or attractive \cite{RonAngevin09} & 
 The feedback remain mostly unadaptive and/or unadapted and researches toward improving this are made difficult by the often few number of participants. & 
 Using learning companions to provide task related feedback and explain to the users how their brain activity is modified when they perform a task.\\
 \hline
Social and emotional Feedback & 
BCI training should be engaging and motivating. & 
Assessing users' states, e.g., the emotional one, often remains unreliable, still needs training and therefore time. & 
Passive BCI could still be used to monitor the learner's state, e.g., emotion or motivation but also the level of attention or fatigue in order to adapt the training. \\ 
 \hline
Cognitive Feedback & 
The feedback should provide insights and guidance to the user. & 
A cognitive model is still lacking and limits the improvement of the training. & 
Using an example based learning companion, which do not require a cognitive model of the task could improve learning. \\ 
 \hline
\end{tabular}
\caption{Summary of the different recommendations, challenges and potential solutions raised in this article.}
\label{tab:summary}
\end{table*}

\subsection{Cognitive feedback}\label{chap:CognitiveFB}
Such as social and emotional feedback, cognitive feedback constitute another challenge and present a great opportunity to improve BCI training. According to Balzer et al. \cite{Balzer89}, providing cognitive feedback \textit{``refers to the process of presenting the person information about the relations in the environment (i.e., task information), relations perceived by the person (i.e., cognitive information), and relations between the environment and the person's perceptions of the environment (i.e., functional validity information)}''. They suggest that the task information is the type of cognitive feedback influencing the most the performance. Therefore, providing BCI users with information about the way they \textit{do} vs. \textit{should} perform the MI-tasks is most likely of the utmost importance.

\subsubsection{BCI Literature}
Currently, the most used cognitive feedback in MI-BCI is the classification accuracy (CA), i.e., the percentage of mental commands that are correctly recognized by the system \cite{Jeunet16e}. While informative, this feedback remains only evaluative: it provides some information about how well the learner \textit{does} perform the task, but no information about how they \textit{should} perform it. Some studies have been led in order to enrich this feedback. \cite{Kaufmann11} proposed a richer ``multimodal'' feedback providing information about the task recognized by the classifier, the strength/confidence in this recognition as well as the dynamics of the classifier output throughout the whole trial. \cite{Sollfrank16} chose to add information concerning the stability of the EEG signals to the standard feedback based on CA, while \cite{Schumacher15} added an explanatory feedback based on the level of muscular relaxation to this CA-based feedback. This additional feedback was used to explain poor CA as a positive correlation had been previously suggested between muscular relaxation and CA. Finally, \cite{Zich15} provided learners with a 2-dimensional feedback based on a basketball metaphor: ball movements along the horizontal axis were determined by classification of contra- versus ipsilateral activity (i.e., between the two brain's hemispheres), whereas vertical movements resulted from classifying contralateral activity of baseline versus MI interval. 

By adding some dimensions to the standard classification CA-based feedback, these feedbacks provided more information to the learner about the way to improve their performance. Nonetheless, all of them are still mainly based on the CA which may not be appropriate to assess users' learning \cite{Lotte17}. Indeed, CA may not reflect properly successful EEG pattern self-regulation. Yet, learning to self-regulate specific EEG patterns, and more specifically to generate stable and distinct patterns for each MI task are the skills to be acquired by the learner \cite{Jeunet17}.

\subsubsection{Learning Companion Literature}
Beside emotional (affective) and social assistance, learning companions can also be designed to provide a cognitive support to the learner. In this perspective, many solutions exist in the field of intelligent tutoring systems (ITS), which use computational tools to tutor the learner. For instance, the companion strategy can be based on the current student learning path compared to an explicit cognitive model of the task which highlights the different solution paths and skills involved \cite{Aleven10}. A learning path gathers the actions taken by the learner (providing an answer, asking for help, taking notes, etc), and the context of these actions (e.g., did the learner attempted an answer before asking for help?). Recognizing learners’ learning path and skills used can also be done using a constraint-based model of the task \cite{Mitrovic10} or a model of the task learnt using relevant machine learning or data mining techniques. Whatever approach is used, the goal is to create a model where a learning companion can act and track learners' actions or behavior to determine how they learn and provide them with an effective cognitive accompaniment or assistance. 

On the sidelines of these cognitive tutors, \textit{example tracing tutors} \cite{Koedinger09} have been newly developed. They elaborate their feedback by comparing the actual strategy of the user with some previous correct and incorrect strategies, which means that they do not require any preexisting cognitive model of the task. This type of tutoring is based on imitating the successful behavior of others. Two types of imitations are possible, one by studying worked examples, the other by directly observing someone else performing the task \cite{VanGog10}.

\subsubsection{Futures challenges}
The latter second type of imitation based training has already proven useful in BCI by \cite{Kondo15}. They showed that BCI training could be enhanced by having users watch someone performing the motor task they imagined. Though providing the users with worked examples has never been tried and might be worth exploring by using a learning companion to provide those worked examples. In order to do so, the users would have to explicit the different strategies they used to control the BCI. One way to do so could be by teaching the companion. This represents a challenge because of the variety of strategies users can use which would then have to be analyzed, but also because the verbalization of motor-related strategies is subjective. Methods developed for clarifying interview and user experience assessment could be adapted in order to clarify these verbalizations \cite{wilson13}. Such researches could be linked to the semiotic training suggested for BCI, which consists in training participants to improve their capacity to associate their mental imagery strategies with their BCI performances \cite{Timofeeva16}. The benefit of these methods is that they do not require a cognitive model of the task though they could help determine learning paths and prove useful to develop such cognitive model. 

Indeed, in order to be able to provide more relevant cognitive feedback to BCI learners, we should first deepen our theoretical knowledge about the MI-BCI skills and about their underlying processes. Very little work has been performed by the community to model MI-BCI tasks and thus the skills to be acquired. Thus, the challenges to address (see \textit{\cref{tab:summary}}) are the following:
\begin{enumerate}
\item Define and implement a computational cognitive model of MI-BCI tasks \cite{Jeunet17}
\item Based on this model, determine which skills should be acquired
\item Based on these skills, define relevant measures of performance
\item Based on these measures of performance, design cognitive feedback to help BCI learner to achieve a high performance, i.e., to acquire the target skills 
\end{enumerate}

\section{DISCUSSION \& CONCLUSION}

In this article, we have shown that BCIs are promising interaction systems enabling users to interact using their brain activity only. However, they require users to train so they can control them, and so far, this training has been suboptimal. Here, we hope we demonstrated how artificial learning companions could contribute to improving this training. In particular, we reviewed how such companions could be used to provide user-adapted and adaptive feedback at the social, emotional and cognitive levels. While there have been various researches on the appearance of BCI feedback, there is almost no research on social, emotional and cognitive feedback for BCI. Learning companion could bridge that gap. Reviewing the literature in learning companion, we suggested various ways to make that happen and the corresponding research challenges that will need to be solved. They are summarized in Table \ref{tab:summary}.

To conclude, the definition from \cite{Chou03} (i.e., ``a learning companion is a computer-simulated character, which has human-like characteristics and plays a non-authoritative role in a social learning environment'') is especially interesting because it involves an exchange of knowledge between the learner and the learning companion. This builds on the idea that, on the one hand the BCI trainee could benefit from social, emotional and cognitive feedback that the learning companion would provide. While on the other hand, the model maintained by the learning companion could benefit from the learners' feedback to be better adapted. 

Both the psychological profile and the cognitive state of the learner have an influence on the capacity to use a BCI and the type of learning companion that can be the most effective. Therefore, creating models to understand 1) which state the users go through while learning 2) how the psychological characteristics and cognitive states of the user influence the learning and finally 3) how to provide an adapted feedback according to the previous points represent a common goal for the BCI and Learning companion fields where both could benefit each other.

Even though we focused on the improvement learning companions could bring to the feedback, the benefits are not limited to it. For example, they could also be used to assess or limit the potential experimenter bias, which occurs when experimenters' expectation or knowledge involuntarily influence their subjects \cite{Rosnow97}. Indeed, they could limit the need for an experimenter and make it easier to perform double blind experiments, where both subjects and experimenters do not know to which experimental group they belong.

\subsubsection*{Acknowledgements}
This work was supported by the French National Research Agency (project REBEL, grant ANR-15-CE23-0013-01) and the European Research Council (project BrainConquest, grant ERC-2016-STG-714567).

\bibliographystyle{abbrv}
\bibliography{WACAI2018}

\begin{thebibliography}{10}

\bibitem{Aleven10}
V.~Aleven, I.~Roll, B.~M. McLaren, and K.~R. Koedinger.
\newblock Automated, unobtrusive, action-by-action assessment of
  self-regulation during learning with an intelligent tutoring system.
\newblock {\em Educational Psychologist}, 45(4):224--233, 2010.

\bibitem{Ang15}
K.~Ang and C.~Guan.
\newblock Brain--computer interface for neurorehabilitation of upper limb after
  stroke.
\newblock {\em Proceedings of the IEEE}, 103(6):944--953, 2015.

\bibitem{Azevedo95}
R.~Azevedo and R.~Bernard.
\newblock A meta-analysis of the effects of feedback in computer-based
  instruction.
\newblock {\em J Educ Comp Res}, 13(2):111--127, 1995.

\bibitem{Balzer89}
W.~Balzer, M.~Doherty, et~al.
\newblock Effects of cognitive feedback on performance.
\newblock {\em Psychological bulletin}, 106(3):410, 1989.

\bibitem{Bangert91}
R.~Bangert-Drowns, C.~Kulik, J.~Kulik, and M.~Morgan.
\newblock The instructional effect of feedback in test-like events.
\newblock {\em Review of educational research}, 61(2):213--238, 1991.

\bibitem{Beale09}
R.~Beale and C.~Creed.
\newblock Affective interaction: How emotional agents affect users.
\newblock {\em International Journal of Human-Computer Studies},
  67(9):755--776, 2009.

\bibitem{Bent17}
O.~Bent, P.~Dey, K.~Weldemariam, and M.~Mohania.
\newblock Modeling user behavior data in systems of engagement.
\newblock {\em Futur Gen Comp Sys}, 68:456--464, 2017.

\bibitem{Bonnet13}
L.~Bonnet, F.~Lotte, and A.~Lécuyer.
\newblock Two brains, one game: design and evaluation of a multiuser {BCI}
  video game based on motor imagery.
\newblock {\em IEEE Transactions on Computational Intelligence and AI in
  games}, 5(2):185--198, 2013.

\bibitem{Burleson07}
W.~Burleson and R.~Picard.
\newblock Gender-specific approaches to developing emotionally intelligent
  learning companions.
\newblock {\em IEEE Intelligent Systems}, 22(4), 2007.

\bibitem{Cabada12}
R.~Cabada, M.~Estrada, C.~Garcia, Y.~P{\'e}rez, et~al.
\newblock Fermat: merging affective tutoring systems with learning social
  networks.
\newblock In {\em Proc ICALT}, pages 337--339, 2012.

\bibitem{Chou03}
C.~Chou, T.~Chan, and C.~Lin.
\newblock Redefining the learning companion: the past, present, and future of
  educational agents.
\newblock {\em Computers \& Education}, 40(3), 2003.

\bibitem{Churamani17}
N.~Churamani, M.~Kerzel, E.~Strahl, P.~Barros, and S.~Wermter.
\newblock Teaching emotion expressions to a human companion robot using deep
  neural architectures.
\newblock In {\em Proc IJCNN}, pages 627--634, 2017.

\bibitem{Clerc16-v1}
M.~Clerc, L.~Bougrain, and F.~Lotte.
\newblock {\em Brain-Computer Interfaces 1: Foundations and Methods}.
\newblock ISTE-Wiley, 2016.

\bibitem{Clerc16-v2}
M.~Clerc, L.~Bougrain, and F.~Lotte.
\newblock {\em Brain-Computer Interfaces 2: Technology and Applications}.
\newblock ISTE-Wiley, 2016.

\bibitem{Duffy03}
B.~Duffy.
\newblock Anthropomorphism and the social robot.
\newblock {\em Robotics and autonomous systems}, 42(3):177--190, 2003.

\bibitem{Dunbar07}
R.~I. Dunbar and S.~Shultz.
\newblock Evolution in the social brain.
\newblock {\em science}, 317(5843), 2007.

\bibitem{Edlinger11}
G.~Edlinger and C.~Guger.
\newblock {\em Social environments, mixed communication and goal-oriented
  control application using a brain-computer interface}, volume 6766 LNCS of
  {\em Lecture {Notes} in {Computer} {Science}}.
\newblock 2011.

\bibitem{Frey14}
J.~Frey, R.~Gervais, S.~Fleck, F.~Lotte, and M.~Hachet.
\newblock Teegi: {T}angible {EEG} interface.
\newblock In {\em Proc ACM UIST}, pages 301--308, 2014.

\bibitem{Goebel04}
R.~Goebel, B.~Sorger, J.~Kaiser, N.~Birbaumer, and N.~Weiskopf.
\newblock Bold brain pong: Self regulation of local brain activity during
  synchronously scanned, interacting subjects.
\newblock In {\em 34th Annual Meeting of the Society for Neuroscience}, 2004.

\bibitem{Goleman95}
D.~Goleman.
\newblock {\em Emotional {Intelligence}. {New} {York}: {Brockman}}.
\newblock Inc, 1995.

\bibitem{Gordon16}
G.~Gordon, S.~Spaulding, J.~Westlund, J.~Lee, L.~Plummer, M.~Martinez, M.~Das,
  and C.~Breazeal.
\newblock Affective personalization of a social robot tutor for children's
  second language skills.
\newblock In {\em AAAI}, pages 3951--3957, 2016.

\bibitem{Hattie07}
J.~Hattie and H.~Timperley.
\newblock The power of feedback.
\newblock {\em Review of educational research}, 77(1):81--112, 2007.

\bibitem{Hornecker11}
E.~Hornecker.
\newblock The role of physicality in tangible and embodied interactions.
\newblock {\em Interactions}, 18(2):19--23, 2011.

\bibitem{Isen01}
A.~Isen.
\newblock An influence of positive affect on decision making in complex
  situations: Theoretical issues with practical implications.
\newblock {\em Journal of consumer psychology}, 11(2):75--85, 2001.

\bibitem{Izuma08}
K.~Izuma, D.~Saito, and N.~Sadato.
\newblock Processing of {Social} and {Monetary} {Rewards} in the {Human}
  {Striatum}.
\newblock {\em Neuron}, 58(2):284--294, Apr. 2008.

\bibitem{Jeunet16}
C.~Jeunet, E.~Jahanpour, and F.~Lotte.
\newblock Why standard brain-computer interface (bci) training protocols should
  be changed: an experimental study.
\newblock {\em Journal of neural engineering}, 13(3):036024, 2016.

\bibitem{Jeunet16e}
C.~Jeunet, F.~Lotte, and B.~N'Kaoua.
\newblock {\em Human Learning for Brain--Computer Interfaces}, pages 233--250.
\newblock Wiley Online Library, 2016.

\bibitem{Jeunet17}
C.~Jeunet, B.~N'Kaoua, and F.~Lotte.
\newblock Towards a cognitive model of {MI-BCI} user training.
\newblock 2017.

\bibitem{Jeunet15a}
C.~Jeunet, C.~Vi, D.~Spelmezan, B.~N’Kaoua, F.~Lotte, and S.~Subramanian.
\newblock Continuous tactile feedback for motor-imagery based brain-computer
  interaction in a multitasking context.
\newblock In {\em Human-Computer Interaction}, pages 488--505, 2015.

\bibitem{Johnson09}
D.~Johnson and R.~Johnson.
\newblock An educational psychology success story: {Social} interdependence
  theory and cooperative learning.
\newblock {\em Educational researcher}, 2009.

\bibitem{Johnson04}
W.~Johnson and P.~Rizzo.
\newblock Politeness in tutoring dialogs:“run the factory, that’s what
  i’d do”.
\newblock In {\em Intelligent Tutoring Systems}, pages 206--243. Springer,
  2004.

\bibitem{Kaufmann11}
T.~Kaufmann, J.~Williamson, E.~Hammer, R.~Murray-Smith, and A.~K{\"u}bler.
\newblock Visually multimodal vs. classic unimodal feedback approach for
  smr-bcis: a comparison study.
\newblock {\em Int. J. Bioelectromagn}, 13:80--81, 2011.

\bibitem{Keller08}
J.~Keller.
\newblock An integrative theory of motivation, volition, and performance.
\newblock {\em Technology, Instruction, Cognition, and Learning}, 6(2):79--104,
  2008.

\bibitem{Kim05}
Y.~Kim.
\newblock Pedagogical agents as learning companions: Building social relations
  with learners.
\newblock In {\em AIED}, pages 362--369, 2005.

\bibitem{Kleih10}
S.~Kleih, F.~Nijboer, S.~Halder, and A.~K{\"u}bler.
\newblock Motivation modulates the {P300} amplitude during brain--computer
  interface use.
\newblock {\em Clinical Neurophysiology}, 2010.

\bibitem{Koedinger09}
K.~Koedinger, V.~Aleven, B.~McLaren, and J.~Sewall.
\newblock Example-tracing tutors: A new paradigm for intelligent tutoring
  systems.
\newblock {\em Authoring Intelligent Tutoring Systems}, pages 105--154, 2009.

\bibitem{Kondo15}
T.~Kondo, M.~Saeki, Y.~Hayashi, K.~Nakayashiki, and Y.~Takata.
\newblock Effect of instructive visual stimuli on neurofeedback training for
  motor imagery-based brain--computer interface.
\newblock {\em Human movement science}, 43:239--249, 2015.

\bibitem{Krach08}
S.~Krach, F.~Hegel, B.~Wrede, G.~Sagerer, F.~Binkofski, and T.~Kircher.
\newblock Can machines think? interaction and perspective taking with robots
  investigated via fmri.
\newblock {\em PloS one}, 3(7):e2597, 2008.

\bibitem{Kubler01b}
A.~K{\"u}bler, N.~Neumann, J.~Kaiser, B.~Kotchoubey, T.~Hinterberger, and
  N.~Birbaumer.
\newblock Brain-computer communication: self-regulation of slow cortical
  potentials for verbal communication.
\newblock {\em Arch phys med rehab}, 82(11), 2001.

\bibitem{Kubler01a}
A.~Kübler, B.~Kotchoubey, J.~Kaiser, J.~Wolpaw, and N.~Birbaumer.
\newblock Brain–computer communication: {Unlocking} the locked in.
\newblock {\em Psychological bulletin}, 127(3):358, 2001a.

\bibitem{Lecuyer08}
A.~L{\'e}cuyer, F.~Lotte, R.~Reilly, R.~Leeb, M.~Hirose, and M.~Slater.
\newblock Brain-computer interfaces, virtual reality, and videogames.
\newblock {\em Computer}, 41(10), 2008.

\bibitem{LeDoux95}
J.~LeDoux.
\newblock Emotion: Clues from the brain.
\newblock {\em Ann rev psych}, 46(1):209--235, 1995.

\bibitem{Leeb07}
R.~Leeb, F.~Lee, C.~Keinrath, R.~Scherer, H.~Bischof, and G.~Pfurtscheller.
\newblock Brain–computer communication: motivation, aim, and impact of
  exploring a virtual apartment.
\newblock {\em IEEE Trans Neur Sys Rehab}, 15(4):473--482, 2007.

\bibitem{Lester97}
J.~Lester, S.~Converse, S.~Kahler, S.~Barlow, B.~Stone, and R.~Bhogal.
\newblock The persona effect: affective impact of animated pedagogical agents.
\newblock In {\em Proc ACM CHI}, 1997.

\bibitem{LotteHDR2016}
F.~Lotte.
\newblock {\em Towards Usable Electroencephalography-based Brain-Computer
  Interfaces}.
\newblock Habilitation thesis ({HDR}), Univ. Bordeaux, 2016.

\bibitem{Lotte15a}
F.~Lotte and C.~Jeunet.
\newblock Towards improved {BCI} based on human learning principles.
\newblock In {\em 3rd International Brain-Computer Interfaces Winter
  Conference}, 2015.

\bibitem{Lotte17}
F.~Lotte and C.~Jeunet.
\newblock Online classification accuracy is a poor metric to study mental
  imagery-based {BCI} user learning: an experimental demonstration and new
  metrics.
\newblock In {\em 7th International BCI Conference}, 2017.

\bibitem{Lotte13}
F.~Lotte, F.~Larrue, and C.~M{\"u}hl.
\newblock Flaws in current human training protocols for spontaneous
  brain-computer interfaces: lessons learned from instructional design.
\newblock {\em Frontiers in human neuroscience}, 7, 2013.

\bibitem{Mathiak15}
K.~Mathiak, E.~Alawi, Y.~Koush, M.~Dyck, J.~Cordes, T.~Gaber, F.~Zepf,
  N.~Palomero-Gallagher, P.~Sarkheil, S.~Bergert, M.~Zvyagintsev, and
  K.~Mathiak.
\newblock Social reward improves the voluntary control over localized brain
  activity in {fMRI}-based neurofeedback training.
\newblock {\em Frontiers in Behavioral Neuroscience}, 9(June), 2015.

\bibitem{Mattout12}
J.~Mattout.
\newblock Brain-{Computer} {Interfaces}: {A} {Neuroscience} {Paradigm} of
  {Social} {Interaction}? {A} {Matter} of {Perspective}.
\newblock {\em Frontiers in Human Neuroscience}, 6, 2012.

\bibitem{Mcquiggan10}
S.~McQuiggan, J.~Robison, and J.~Lester.
\newblock Affective transitions in narrative-centered learning environments.
\newblock {\em Educational Technology \& Society}, 13(1):40--53, 2010.

\bibitem{Mercier14}
J.~Mercier-Ganady, F.~Lotte, E.~Loup-Escande, and A.~Marchal, M.and~Lecuyer.
\newblock The mind-mirror: See your brain in action in your head using eeg and
  augmented reality.
\newblock In {\em Virtual Reality (VR), 2014 iEEE}, pages 33--38. IEEE, 2014.

\bibitem{Merrill07}
M.~Merrill.
\newblock First principles of instruction: a synthesis.
\newblock {\em Trends and issues in instructional design and technology},
  2:62--71, 2007.

\bibitem{Millan10}
J.~Mill\'an, R.~Rupp, G.~M\"uller-Putz, R.~Murray-Smith, C.~Giugliemma,
  M.~Tangermann, C.~Vidaurre, F.~Cincotti, A.~K\"ubler, R.~Leeb, C.~Neuper,
  K.-R. M\"uller, and D.~Mattia.
\newblock Combining brain-computer interfaces and assistive technologies:
  State-of-the-art and challenges.
\newblock {\em Frontiers in Neuroprosthetics}, 2010.

\bibitem{Mitrovic10}
A.~Mitrovic.
\newblock Modeling domains and students with constraint-based modeling.
\newblock {\em Advances in intelligent tutoring systems}, pages 63--80, 2010.

\bibitem{Muhl14}
C.~M{\"u}hl, B.~Allison, A.~Nijholt, and G.~Chanel.
\newblock A survey of affective brain computer interfaces: principles,
  state-of-the-art, and challenges.
\newblock {\em Brain-Computer Interfaces}, 1(2):66--84, 2014.

\bibitem{Narciss04}
S.~Narciss and K.~Huth.
\newblock How to design informative tutoring feedback for multimedia learning.
\newblock {\em Instructional design for multimedia learning}, 181195, 2004.

\bibitem{Neuper10}
C.~Neuper and G.~Pfurtscheller.
\newblock {\em Brain-Computer Interfaces}, chapter Neurofeedback Training for
  {BCI} Control, pages 65--78.
\newblock The Frontiers Collection, 2010.

\bibitem{Nijboer08}
F.~Nijboer, A.~Furdea, I.~Gunst, J.~Mellinger, D.~McFarland, N.~Birbaumer, and
  A.~Kübler.
\newblock An auditory brain–computer interface ({BCI}).
\newblock {\em J Neur Meth}, 2008.

\bibitem{Norman94}
D.~Norman.
\newblock How might people interact with agents.
\newblock {\em Comm ACM}, 37(7), 1994.

\bibitem{Obbink12}
M.~Obbink, H.~Gürkök, D.~Plass-Oude~Bos, G.~Hakvoort, M.~Poel, and
  A.~Nijholt.
\newblock {\em Social interaction in a cooperative brain-computer interface
  game}.
\newblock LNICST. 2012.

\bibitem{Pfurtscheller01}
G.~Pfurtscheller and C.~Neuper.
\newblock Motor imagery and direct brain-computer communication.
\newblock {\em proceedings of the IEEE}, 89(7):1123--1134, 2001.

\bibitem{Pillette17}
L.~Pillette, C.~Jeunet, B.~Mansencal, R.~N'Kambou, B.~N'Kaoua, and F.~Lotte.
\newblock Peanut: Personalised emotional agent for neurotechnology
  user-training.
\newblock In {\em 7th International BCI Conference}, 2017.

\bibitem{Ramos12}
A.~Ramos-Murguialday, M.~Sch{\"u}rholz, V.~Caggiano, M.~Wildgruber, A.~Caria,
  E.~Hammer, S.~Halder, and N.~Birbaumer.
\newblock Proprioceptive feedback and brain computer interface (bci) based
  neuroprostheses.
\newblock {\em PloS one}, 7(10):e47048, 2012.

\bibitem{Robinson02}
M.~Robinson and G.~Clore.
\newblock Belief and feeling: evidence for an accessibility model of emotional
  self-report.
\newblock {\em Psychological bulletin}, 128(6):934, 2002.

\bibitem{RonAngevin09}
R.~Ron-Angevin and A.~D{\'\i}az-Estrella.
\newblock Brain--computer interface: Changes in performance using virtual
  reality techniques.
\newblock {\em Neur let}, 449(2):123--127, 2009.

\bibitem{Rosnow97}
R.~Rosnow and R.~Rosenthal.
\newblock {\em People studying people: Artifacts and ethics in behavioral
  research}.
\newblock WH Freeman, 1997.

\bibitem{Ryan00}
R.~Ryan and E.~Deci.
\newblock Self-determination theory and the facilitation of intrinsic
  motivation, social development, and well-being.
\newblock {\em Am psych}, 55(1):68, 2000.

\bibitem{Saerbeck10}
M.~Saerbeck, T.~Schut, C.~Bartneck, and M.~Janse.
\newblock Expressive robots in education: varying the degree of social
  supportive behavior of a robotic tutor.
\newblock In {\em Proc CHI}, pages 1613--1622, 2010.

\bibitem{Schmitz10}
M.~Schmitz.
\newblock Tangible interaction with anthropomorphic smart objects in
  instrumented environments.
\newblock 2010.

\bibitem{Schumacher15}
J.~Schumacher, C.~Jeunet, and F.~Lotte.
\newblock Towards explanatory feedback for user training in brain-computer
  interfaces.
\newblock In {\em Proc IEEE SMC}, pages 3169--3174, 2015.

\bibitem{Sepulveda16}
P.~Sepulveda, R.~Sitaram, M.~Rana, C.~Montalba, C.~Tejos, and S.~Ruiz.
\newblock How feedback, motor imagery, and reward influence brain
  self-regulation using real-time fmri.
\newblock {\em Human brain mapping}, 37(9):3153--3171, 2016.

\bibitem{Sexton15}
C.~Sexton.
\newblock The overlooked potential for social factors to improve effectiveness
  of brain-computer interfaces.
\newblock {\em Frontiers in Systems Neuroscience}, 9(May):1--5, 2015.

\bibitem{Shute08}
V.~Shute.
\newblock Focus on formative feedback.
\newblock {\em Rev Edu Res}, 78:153--189, 2008.

\bibitem{Sitaram16}
R.~Sitaram, T.~Ros, L.~Stoeckel, S.~Haller, F.~Scharnowski, J.~Lewis-Peacock,
  N.~Weiskopf, M.~Blefari, M.~Rana, E.~Oblak, et~al.
\newblock Closed-loop brain training: the science of neurofeedback.
\newblock {\em Nature Reviews Neuroscience}, 2016.

\bibitem{Sollfrank16}
T.~Sollfrank, A.~Ramsay, S.~Perdikis, J.~Williamson, R.~Murray-Smith, R.~Leeb,
  J.~Mill{\'a}n, and A.~K{\"u}bler.
\newblock The effect of multimodal and enriched feedback on {SMR}-{BCI}
  performance.
\newblock {\em Clinical Neurophysiology}, 127(1):490--498, 2016.

\bibitem{Terzis12}
V.~Terzis, C.~Moridis, and A.~Economides.
\newblock The effect of emotional feedback on behavioral intention to use
  computer based assessment.
\newblock {\em Computers \& Education}, 59(2):710--721, 2012.

\bibitem{Timofeeva16}
M.~Timofeeva.
\newblock Semiotic training for brain-computer interfaces.
\newblock In {\em Proc FedCSIS}, pages 921--925, 2016.

\bibitem{vanErp12}
J.~van Erp, F.~Lotte, and M.~Tangermann.
\newblock Brain-computer interfaces: Beyond medical applications.
\newblock {\em IEEE Computer}, 45(4):26--34, 2012.

\bibitem{VanGog10}
T.~Van~Gog and N.~Rummel.
\newblock Example-based learning: Integrating cognitive and social-cognitive
  research perspectives.
\newblock {\em Edu Psych Rev}, 22(2):155--174, 2010.

\bibitem{Williams97}
S.~Williams.
\newblock Teachers' written comments and students' responses: A socially
  constructed interaction.
\newblock 1997.

\bibitem{wilson13}
C.~Wilson.
\newblock {\em Interview techniques for UX practitioners: A user-centered
  design method}.
\newblock Newnes, 2013.

\bibitem{Witte13}
M.~Witte, S.~Kober, M.~Ninaus, C.~Neuper, and G.~Wood.
\newblock Control beliefs can predict the ability to up-regulate sensorimotor
  rhythm during neurofeedback training.
\newblock {\em Frontiers in Human Neuroscience}, 7, 2013.

\bibitem{Ybarra08}
O.~Ybarra, E.~Burnstein, P.~Winkielman, M.~Keller, M.~Manis, E.~Chan, and
  J.~Rodriguez.
\newblock Mental exercising through simple socializing: Social interaction
  promotes general cognitive functioning.
\newblock {\em Personality and Social Psychology Bulletin}, 34(2):248--259,
  2008.

\bibitem{Zander09}
T.~Zander and S.~Jatzev.
\newblock Detecting affective covert user states with passive brain-computer
  interfaces.
\newblock In {\em Proc ACII}, pages 1--9, 2009.

\bibitem{Zich15}
C.~Zich, S.~Debener, M.~De~Vos, S.~Frerichs, S.~Maurer, and C.~Kranczioch.
\newblock Lateralization patterns of covert but not overt movements change with
  age: An eeg neurofeedback study.
\newblock {\em Neuroimage}, 116:80--91, 2015.

\end{thebibliography}

\end{document}